# Graphene-based Solitons for Spatial Division Multiplexed Switching

Jonathan K. George[1], Volker J. Sorger[1*]

[1]Department of Electrical and Computer Engineering, The George Washington University, Washington, D.C. 20052, USA

*Corresponding author: sorger@gwu.edu

**Abstract:**

**Spatial division multiplexing utilizes the directionality of light's propagating k-vector to separate it into distinct spatial directions. Here we show that the anisotropy of orthogonal spatial Solitons propagating in a single Graphene monolayer results in phase-based multiplexing. We use the self-confinement properties of spatial Solitons to increase the usable density of states (DOS) of this switching-system. Furthermore, we show that crossing two orthogonal Solitons exhibit a low (0.035 dB) mutual disturbance from another enabling independent k-vector switching. The efficient utilization of the DOS and multiplexing in real-space enables data processing parallelism with applications in optical networking and computing.**

Main Body:

In contrast to light in linear media, in highly non-linear media is able to modify the index of refraction of an electromagnetic wave along its own path, enabling the formation of spatial Solitons [1]. A spatial Soliton changes the index of the material just enough to balance spatial dispersion, propagating with an unchanging mode profile. This spatial stability enables light redirection and routing applications [2-4]. Moreover non-linear field interactions of Solitons with photons give rise to all-optical switching and beam-steering applications [5]. While a steered output of a single signal is useful for many applications, performance in computing and data processing strongly depends on the switching density of multiple signals. Here we show phase-based beam steering of spatial Solitons and demonstrate independent beam control of two pairs of crossing in-plane spatial Solitons. Such independence of two orthogonal pairs of all-optical switches enables applications towards all-optical computing.

Photonic waveguides are the basis for monolithic integrated photonic devices [6-9]. The shape and area of the waveguide cross-section defines the modes that are able to propagate in the waveguide, and the cross-section of the waveguide is a directionally dependent property of the waveguide [10-12]. When waveguides cross each other orthogonally, some amount of light from one waveguide will cross couple to the other [13]. Such crosstalk is due to the nature of the modes traveling within the orthogonal waveguides; at the waveguide-crossing neither waveguide can support its primary mode. Thus, the electromagnetic waves experience an impedance mismatch between the regions of the intersection versus the waveguide portion. At this junction, light begins to diffract away from the intersection due to scattering, diffraction, and modal mismatches intersection cross coupling insertion loss of about 0.16 dB in silicon photonics [13]. In cases where the waveguides support primary modes of differing wavelengths, the respective insertion loss of each waveguide increases, since only certain spectral regions can be optimized for cross-coupling due to dispersion.

The mathematical basis of this problem lies in the isotropicity of the refractive index of the two crossing waveguides. Due to the isotropic nature of the index, the two directions of propagation cannot be separated. In an anisotropic media this is not the case. Here, each direction of the electric field can experience a different index and two beams propagating with distinct directions will not necessarily experience the same index profile. Spatial Solitons must be anisotropic due to the field vector-dependence of the nonlinear media creating them. This gives rise to the unique ability to cross each other orthogonally which can be realized with Solitons as discuss here, leading to little mutual interaction useful to maximize switching density.

Solitons occur in non-linear physical systems, including nonlinear optical systems, and can be classified as temporal or spatial. Optical Solitons have been investigated for many applications in both communications and computing [5, 14-19]. Temporal Solitons increases the DOS relative to linear systems by holding pulses of light together in time, while spatial Solitons increases the DOS relative to linear systems by acting as its own waveguide, holding light together in space. In this sense the spatial Soliton maintains constant width, i.e. beam divergence, while the temporal Soliton maintains constant pulse shape, i.e. temporal dispersion. For Kerr nonlinear media, such as Graphene, in which the refractive index is a function of intensity, secant spatial Solitons are given by the nonlinear Schrödinger equation [20]

$$i\frac{\partial}{\partial t}u + \frac{\partial^2}{\partial x^2}u + g|u|^2 u = 0$$

$$t = 2ky \quad (1)$$

Where $u$ is the envelope of the electric field, with a solution Eq. (2) that describes optical spatial Solitons propagating in a 2-dimensional plane with a secant intensity profile [21].

$$E(x,z) = \frac{1}{ka_0}\sqrt{\frac{n_0}{n_2}}\exp\left(\frac{iz}{2ka_0^2}\right)\text{sech}\left(\frac{x - x_0}{a_0}\right) \quad (2)$$

where $k$ is the wavevector, $n_0$ and $n_2$ are the indices of the nonlinear material independent of field strength and dependent on field strength respectively, $n = n_0 + n_2 E^2$. Graphene is a Kerr-type nonlinear optical material with an index of refraction that increases with beam intensity originating from its linear gapless energy dispersion of charge carriers in Graphene that relate energy $E$ to momentum $p$ with $E(p) \propto \pm|p|$ and velocity $v(t) \propto p(t)/|p(t)|$ rather than the parabolic dispersion equation, $E(p) \propto \pm p^2$, and velocity $v(t) \propto p(t)$ [22]. The non-linear term in the carrier velocity creates non-linear harmonics resulting in third order susceptibility, i.e. Kerr nonlinearity. When the nonlinear susceptibility is applied to the electric field in vector form, the $c_{gr}^{(3)}$ is a fourth-rank tensor. This susceptibility has been measured experimentally to be $\left|c_{gr}^{(3)}\right| \approx 1.5 \times 10^{-7} [esu] \approx 2.09 \times 10^{-15} [m^2 V^{-2}]$ [23]. Then the nonlinear polarization, $P$, is a vector, composed of the set of susceptibility tensors of increasing order acting on the electric field, $P = \epsilon_0 \left( \chi^{(1)} E + \chi^{(2)} EE + \chi^{(3)} EEE + ... \right)$ [24]. Assuming the susceptibility is isotropic, from the vector nature of the electric field, it follows that the index of refraction for a wave with the electric field propagating inside a Kerr-type media such as Graphene is anisotropic even when the linear refractive index of the material is isotropic. This results in an anisotropically-graduated index of refraction that varies proportionally with intensity in the direction of electric field [25],

$$n = n_0 + n_2 I = n_0 + \frac{3}{2 n_0^2 \epsilon_0 c} \chi^{(3)} \frac{1}{2} n_0 \epsilon_0 c |E_x|^2 \qquad (3)$$

then for Graphene, $n \approx 3 + 5.225 \times 10^{-16} |E_x|^2$. From this we see that two optical spatial solitons crossing orthogonally to each other in Graphene will have minimal interaction with another; each with orthogonally directed electric fields will independently affect orthogonally directed changes to the index of refraction. This independence increases the DOS.

Based on these arguments, we are interested in investigating options to optically switch a Soliton. For this we select the switching mechanism in a Graphene-based slab-waveguide, which has been previously shown to support optical spatial Solitons [26] (Fig. 1). Graphene is chosen because it fulfills multiple functions simultaneously; a) it serves as a Soliton-generation material, b) it bears a high intrinsic Kerr non-linearity, and c) allows for nanoscale dimensionality enabling compact designs [10, 12]. We note that the modal overlap of a Graphene is relatively low in a diffraction-limited waveguide, but can approach one percent in plasmonic slot-waveguides [12]. First, we evaluate the cross talk between two orthogonally directed Solitons (Fig. 1(a)) by comparing the transmitted power of a Soliton beam with and without a second intersecting Soliton. The results indeed confirm a minimal interaction between the orthogonally intersecting in-phase Solitons as quantified by a low interaction between the two beams of only 0.035 dB in terms of power at the output. The reason for this is that through the nonlinear Kerr effect, the high power in-plane field induces an anisotropic increase in the electric permittivity, causing self-confinement of the beam in the direction of propagation. We note that while the used beam intensity is high, $2.25 \times 10^7 [V/m]$, it is below the field breakdown voltage of $SiO_2$, $3 \times 10^9 [V/m]$ [27].

We next investigate the interaction of two parallel Solitons propagating at a sub-wavelength distances to each other since we have high information-processing density application in mind (Fig. 1(b)). Here the two beams are directed into the Graphene monolayer from the same edge and in the same direction. We find that the spatial distribution at the output depends on the phase difference between the two input beams (Fig. 2). Effectively, this phase difference either pulls the output towards one side (i.e phase change, Fig. 2(a)) or splits the output into two beams (i.e phase change, Fig. 2(c)). Such strong all-optical interaction is a direct result of the Soliton property in Graphene, and the high wavefunction overlap between the two beams being separated by less than one wavelength from each other.

Next we are interested in evaluating the output's spatial distribution dependence on the relative separation distance of the two entering Soliton beams (Fig. 1(b)) to find the maximum separation between the two output peaks (Fig. 3(a)). This point would represent an optimal separation distance for switching. Our results show that as the beams are separated the interference between the two beams continues to create a phase-dependent spatial distribution in the output power up until the beams are separated by over twice their operating wavelength of 850 nm (Fig. 3(b)). The peak separation at the output without decreasing FWHM is found near 500 nm.

Lastly, we combine both previous concepts to show spatial switching functionality; to achieve this we direct two beam pairs along two orthogonal edges of the Graphene-oxide heterostructure monolayer (Fig. 1(c)). Indeed, we find that the minimal cross talk between the two-Soliton pairs preserve the phase dependent spatial distribution at each output (Fig. 4). The beams were simulated in their four potential states: a single pair in-phase, a single pair out-of-phase, two pairs in-phase, and two pairs out-of-phase (Fig. 4). The results demonstrate the independence of the orthogonal pairs of beams. That is, a pair in one state has minimal impact on the crossing pair irrespectively of the state of either pair. This is made possible by the tensor nature of the third order non-linearity resulting in anisotropic intensity dependent adjustments to the refractive index. Such confinement with crossing independence is impossible to achieve in linear isotropic material systems and points to the unique value of spatial optical solitons in dense information processing systems.

In summary, we have shown that spatial division multiplexing in Graphene waveguides is possible enabled by the independence of orthogonally propagating Solitons. We have contrasted this with the dependence between two parallel propagating Solitons and combined both concepts to a form a two-pair Soliton system with independence in each propagation direction. While the switching performance of spatial Solitons is extraordinarily fast, with deflection occurring at the speed of propagation, practical implementation of this device would require overcoming the challenges of coupling into the thin Graphene layer. The context of this work is relevant for optical computing and signal processing, where the ability of optical signals to be multiplexed within the same physical space becomes dominant over the density of components. Signal multiplexing has already been achieved to a limited extent in optical systems with WDM. Extending the concept of multiplexing to include propagation direction has the potential to create another dimension for footprint reuse in future optical systems. The optimal optical computer will take advantage of all possible forms of signal multiplexing to maximize the number of concurrent computational states within its physical bounds.

**Funding.** Air Force Office of Scientific Research-Young Investigator Program (FA9550-14-1-0215, FA9550-14-1-0378).

**Figures**

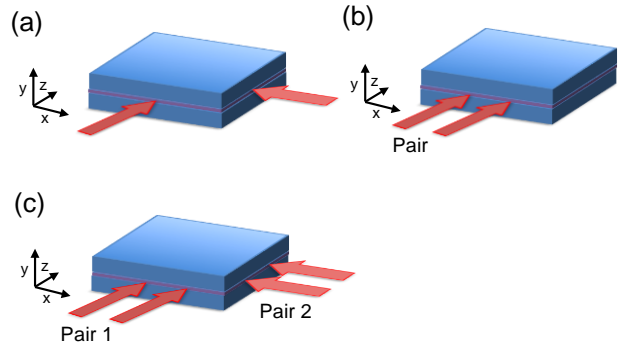

Fig. 1. Schematic and setup of Graphene-based k-vector Soliton switch.. The switch consists of two 300nm $SiO_2$ layers sandwiching a 20nm Graphene monolayer. (a) Two 850nm beams are directed orthogonally in the plane of the Graphene and cross in the center. (b) A pair of 850nm beams is directed with the electric field in plane into the Graphene monolayer to generate self-guided spatial Solitons originating from non-linear Kerr effect. The relative phase of the beams allows controlling the mutual propagation direction of the beams (i.e. k-vector control). (c) Two pairs of 850nm beams are directed into orthogonal edges of the device for testing switching independence of pair 1 from pair 2 to enable dense crossings enabled by the anisotropic index of Graphene and hence the Solitons. All simulations were completed with Lumerical FDTD with a graphene model using a refractive index of 3 without a complex part, and third-order susceptibility of $\left|c_{gr}^{(3)}\right|$ @ $1.5 \times 10^{-7} [esu]$ @ $2.09 \times 10^{-15} [m^2 V^{-2}]$ [23].

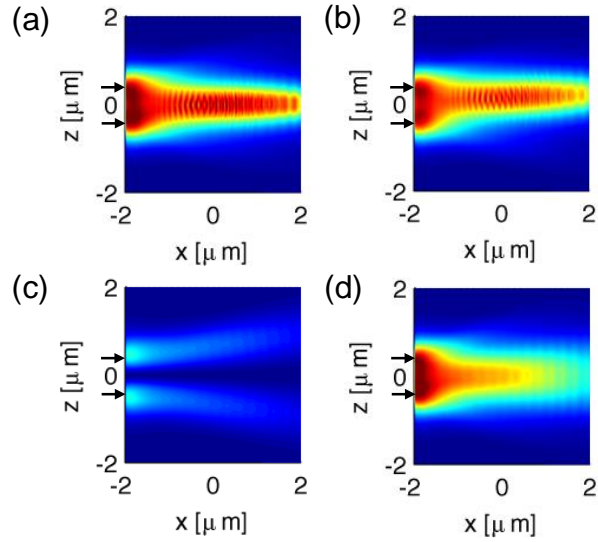

Fig. 2. Phase dependence of two Soliton beams for beam control. Beams propagating from left to right along the x axis (arrows). (a) The power in the plane of the Graphene monolayer when two beams along the z-axis edge are in phase, (b) out of phase by $\pi/4$, (c) out of phase by $\pi$, producing phase-dependent spatial switching at the output. (d) Two beams in phase with each other with Graphene non-linearity disabled shows spatial dispersion.

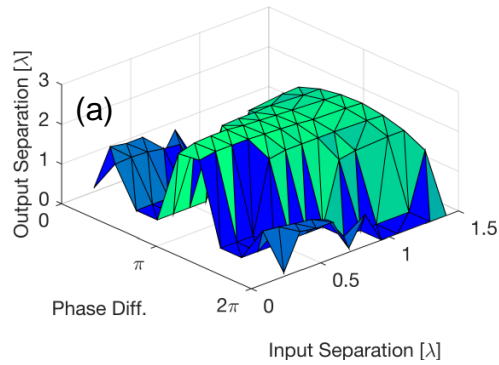

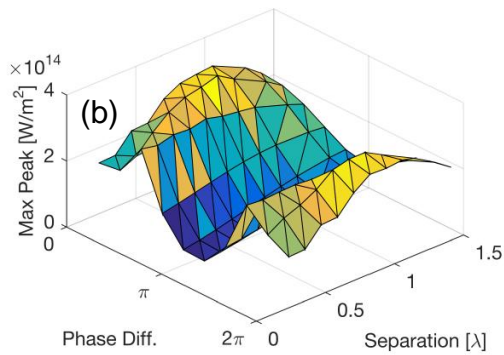

Fig. 3. Soliton beam control sensitivity analysis. (a) A sweep of the output separation as a function of beam phase difference and spatial input beam separation in the configuration of Fig. 1(b) shows that the maximum output separation occurs when the beams are out of phase with sensitivity growing as the beam separation is decreased. (b) A sweep of the largest peak over beam phase difference and input separation also in the configuration of Fig. 1(b) shows an output maximum when the phase difference is minimal and the two beams merge into one

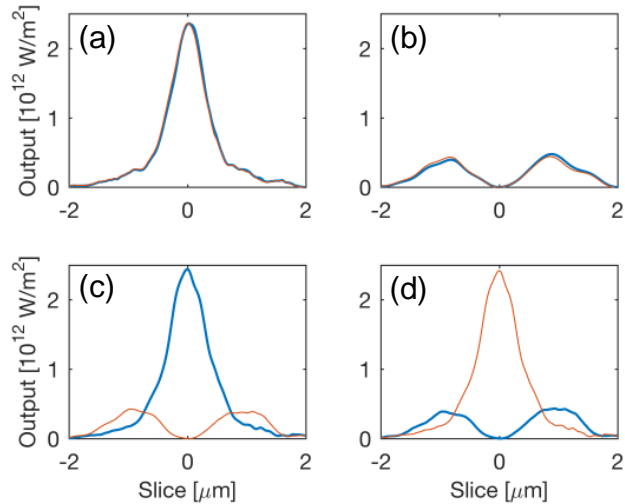

Fig. 4. Crossing Soliton output independence. Output power profile of beam pairs across the x direction (thin line) and orthogonal pair of beams in z direction (thick line) with (a) no phase difference in either pairs (b) π phase difference in z directed pair (c) π phase in x directed pair (d) π phase difference in both pairs.

**References**

1. E. Infeld and G. Rowlands, (Cambridge University Press, 2000).
2. C. Ye, K. Liu, R. A. Soref, and V. J. Sorger, Nanophotonics **4,** 261-268 (2015).
3. S. Sun, A. A. Badawy, V. Narayana, T. El-Ghazawi, and V. J. Sorger, IEEE Photonics Journal **7,** 1-14 (2015).
4. A. Fratalocchi, C. Dodson, R. Zia, P. Genevet, E. Verhagen, H. Altug, and V. Sorger, NATURE NANOTECHNOLOGY **10,** 11-15 (2015).
5. J. Hübner, H. M. van Driel, and J. S. Aitchison, Opt. Lett. **30,** 3168-3170 (2005).
6. C. Huang, R. J. Lamond, S. K. Pickus, Z. R. Li, and V. J. Sorger, IEEE Photonics Journal **5,** 2202411-2202411 (2013).
7. S. K. Pickus, S. Khan, C. Ye, Z. Li, and V. J. Sorger, IEEE Photonic Society, Research highlights **27,** 4-10 (2013).
8. K. Liu, ACS photonics **3,** 233; 233-242; 242 (2016).
9. N. Li, K. Liu, V. Sorger, and D. Sadana, SCIENTIFIC REPORTS **5,** 14067 (2015).
10. C. Ye, S. Khan, Z. Li, E. Simsek, and V. Sorger, IEEE JOURNAL OF SELECTED TOPICS IN QUANTUM ELECTRONICS **20,** (2014).
11. Z. Ma, Z. Li, K. Liu, C. Ye, and V. Sorger, NANOPHOTONICS **4,** 198-213 (2015).
12. Z. Ma, IEEE journal of selected topics in quantum electronics1; 1-1; 1 (2016).
13. W. Bogaerts, P. Dumon, D. Van Thourhout, and R. Baets, Opt. Lett. **32,** 2801-2803 (2007).
14. W. L. Kath, A. Mecozzi, P. Kumar, and C. G. Goedde, Opt. Commun. **157,** 310-326 (1998).
15. R. Schiek, Y. Baek, G. Stegeman, and W. Sohler, Opt. Quant. Electron. **30,** 861-879 (1998).
16. J. K. Jang, M. Erkintalo, S. Coen, and S. G. Murdoch, Nature communications **6,** 7370 (2015).
17. S. Wabnitz, Opt. Lett. **18,** 601 (1993).

18. R. W. Boyd, D. J. Gauthier, and A. L. Gaeta, Opt. Photonics News **17,** 18-23 (2006).
19. W. J. Firth and A. J. Scroggie, Phys. Rev. Lett. **76,** 1623-1626 (1996).
20. R. Y. Chiao, E. Garmire, and C. H. Townes, Phys. Rev. Lett. **13,** 479-482 (1964).
21. J. S. Aitchison, A. M. Weiner, Y. Silberberg, M. K. Oliver, J. L. Jackel, D. E. Leaird, E. M. Vogel, and P. W. Smith, Opt. Lett. **15,** 471-473 (1990).
22. S. Mikhailov, PHYSICAL REVIEW B **90,** (2014).
23. E. Hendry, P. J. Hale, J. Moger, A. K. Savchenko, and S. A. Mikhailov, Phys. Rev. Lett. **105,** 097401 (2010).
24. P. Paufler, Journal of Applied Crystallography **45,** 613-613 (2012).
25. R. W. Boyd, *Nonlinear Optics (Third Edition),* R. W. Boyd, ed. (Academic Press, 2008), pp. 1-67.
26. M. L. Nesterov, J. Bravo-Abad, A. Y. Nikitin, F. J. García-Vidal, and L. Martin-Moreno, Laser & Photonics Reviews **7,** 7-11 (2013).
27. Li-Mo Wang, 2006 25th International Conference on Microelectronics, pp. 576-579.